\documentclass[aps,letterpaper,prl,twocolumn,showpacs,superscriptaddress]{revtex4}
\usepackage{bm,color,amsmath,amssymb,graphicx}
\usepackage[pdftex,
            pdfauthor={Yang-Le Wu},
            pdftitle={Bloch Model Wave Functions and Pseudopotentials for All Fractional Chern Insulators}]{hyperref}
\providecommand{\Wsq}{W_{\protect\rule{3.5pt}{3.5pt}}}

\begin{document}

\title{Bloch Model Wave Functions and Pseudopotentials for All Fractional Chern Insulators}

\author{Yang-Le Wu}
\affiliation{Department of Physics, Princeton University, Princeton, New Jersey 08544, USA}
\author{N. Regnault}
\affiliation{Department of Physics, Princeton University, Princeton, New Jersey 08544, USA}
\affiliation{Laboratoire Pierre Aigrain, ENS and CNRS, 24 rue Lhomond, 75005 Paris, France}
\author{B. Andrei Bernevig}
\affiliation{Department of Physics, Princeton University, Princeton, New Jersey 08544, USA}

\begin{abstract}
We introduce a Bloch-like basis in a $C$-component lowest Landau level 
fractional quantum Hall (FQH) effect,
which entangles the real and internal degrees of freedom and preserves an 
$N_x\times N_y$ full lattice translational symmetry.
We implement the Haldane pseudopotential Hamiltonians in this new basis.
Their ground states are the model FQH wave functions, and our Bloch basis 
allows for a \emph{mutatis mutandis} transcription of these model wave 
functions to the fractional Chern insulator (FCI) of arbitrary Chern number $C$, 
obtaining wave functions different from all previous proposals.
For $C>1$, our wave functions are related to color-dependent magnetic-flux 
inserted versions of Halperin and non-Abelian color-singlet states.
We then provide large-size numerical results for both the $C=1$ and $C=3$ 
cases.
This new approach leads to improved overlaps compared to previous proposals.
We also discuss the adiabatic continuation from the FCI to the FQH in our
Bloch basis, both from the energy and the entanglement spectrum perspectives.
\end{abstract}
\maketitle

Recently, several groups showed that gapped topological phases resembling the 
fractional quantum Hall (FQH) effects can be stabilized in a flat band 
with Chern number $C\neq 0$ by strong electronic interactions in the 
absence of a magnetic field~\cite{Sheng11:FCI,Neupert11:FCI,Regnault11:FCI}.
These are named fractional Chern insulators (FCI).
Most of the research efforts have been focused on the case of $C=1$:
In various lattice
models~\cite{Haldane88:Honeycomb,Sun11:Flatband,Tang11:Kagome,Hu11:Ruby}, 
several groups have provided compelling 
evidence~\cite{Sheng11:FCI,Neupert11:FCI,Regnault11:FCI,Parameswaran12:W-inf,
Bernevig12:Counting,Wu12:Zoology,Wang12:MR,Goerbig12,Roy12:Geometry,
Venderbos12:t2g,Wang11:FCI-Boson,Neupert11:Z2,Neupert12:Conductivity,
Murthy11:CF,Murthy12:CF,Kourtis12:Triangular,Lee12:Wannier,Wu12:Hofstadter}
for the presence of the Read-Rezayi 
series~\cite{Moore91:MR,Read99:RR,Bernevig12:Counting,Wu12:Zoology,Wang12:MR} 
as well as the composite-fermion~\cite{Jain89:CF,Liu12:CF,Lauchli12:Hierarchy} 
FQH states. The correlated phases in 
Chern bands with $C>1$~\cite{Wang11:Dice,Barkeshli12:Nematic,Trescher12:Flatband,
Yang12:Flatband,Wang12:C2}, however, are more intricate.
Numerical studies found both bosonic~\cite{Wang12:C2,Liu12:Higher,Sterdyniak12:Higher} 
and fermionic~\cite{Liu12:Higher,Grushin12:Dispersion} 
topological phases resembling the color $\mathrm{SU}(C)$ version of the 
Halperin~\cite{Halperin83} and the non-Abelian 
spin-singlet~\cite{Ardonne99:Singlet} (NASS) 
states~\cite{Sterdyniak12:Higher}, but with clear deviations~\cite{Sterdyniak12:Higher}.

To understand these novel topological phases, a series of approaches was put 
forward.
For $C=1$, one can identify the nature of these states
(1)~through a folding principle~\cite{Regnault11:FCI,Bernevig12:Counting} that 
links the FCI and FQH quantum numbers,
(2)~through the entanglement spectrum~\cite{Li08:ES,Sterdyniak11:PES} of the ground 
states~\cite{Regnault11:FCI,Wu12:Zoology},
and (3)~through overlaps with model states obtained from replacing the
lowest Landau level (LLL)
orbitals with hybrid Wannier states, but leaving the occupation-number weights 
unchanged~\cite{Qi11:Wavefunction,Barkeshli12:Nematic}.
After proper gauge fixing~\cite{Wu12:Wannier}, high 
overlaps were obtained~\cite{Wu12:Wannier,Scaffidi12:Adiabatic,Liu13:Wannier} 
from the last approach and FCI-FQH adiabatic continuity was 
demonstrated~\cite{Scaffidi12:Adiabatic,Liu13:Wannier}.

For $C>1$, the finite-size numerical results are harder to understand.
The FCI equivalent of the Halperin states was proposed to occur at Abelian 
filling factors~\cite{Barkeshli12:Nematic}. 
The particle entanglement spectrum~\cite{Sterdyniak12:Higher}, however, shows 
a clear discrepancy from such states.
We are also unable to consistently implement the exclusion principle for 
colorful FQH model states~\cite{Estienne12:Singlet,Ardonne11:Squeezing}
in the Wannier basis.
Naively, a $C$-component quantum Hall system contains $C$ decoupled 
copies of the LLL, each having a unity Chern number over a Brillouin zone (BZ) 
consisting of $N_\phi=N_xN_y/C$ momenta~\cite{Bernevig12:Counting}.
This appears to be very different from the single Chern number $C$ 
manifold of the lattice BZ of $N_xN_y$ momenta,
especially when $N_xN_y/C\not\in\mathbb{Z}$.

In this Letter, we break away from previous approaches and
construct in a $C$-component LLL a momentum-space basis that mimics the 
$N_x\times N_y$ Bloch states in the Chern band.
These new one-body basis states entangle the color and 
the real spaces and form a single $N_x\times N_y$ Brillouin zone with 
flat Berry curvature and Chern number $C$, regardless of lattice 
size commensuration with $C$.
This leads to a new mapping between FCI with arbitrary $C$ on a lattice 
of arbitrary size and a $C$-component FQH system.
Our mapping operates directly in Bloch momentum space
and utilizes the full lattice translational symmetry,
which removes the huge computational cost 
of~\cite{Wu12:Wannier,Scaffidi12:Adiabatic}.
For $C=1$, our construction is equivalent to the Wannier 
construction~\cite{Qi11:Wavefunction}, except for a new gauge fixing that 
improves the overlaps (than~\cite{Wu12:Wannier,Liu13:Wannier}).
For $C>1$, our model FCI states are equivalent to a new, color-dependent 
magnetic-flux inserted version of the Halperin or the NASS states, different 
from the existing proposal~\cite{Barkeshli12:Nematic}.
The FCI wave functions produced by our approach have the correct entanglement 
spectrum~\cite{Wu12:Zoology,Sterdyniak12:Higher}.
We demonstrate large overlaps for previously unattained sizes between our 
model FCI wave functions and numerics for both $C=1$ and the uncharted case of 
$C>1$.

Consider a translationally invariant two-dimensional (2D) band insulator on an 
$N_x\times N_y$ lattice with $N_o$ orbitals per unit cell indexed by $b$.
The Bravais lattice is $m_x\mathbf{b}_x+m_y\mathbf{b}_y$, with 
$(m_x,m_y)\in\mathbb{Z}^2$ and the primitive translation vectors $\mathbf{b}_x$ and 
$\mathbf{b}_y$.
We focus on a single Chern band of Bloch states $|\mathbf{k}\rangle$, labeled 
by momentum $\mathbf{k}=\sum_\alpha k_\alpha\mathbf{g}_\alpha$, with 
$k_\alpha\in\mathbb{Z}$ and 
$\mathbf{g}_\alpha\cdot\mathbf{b}_\beta=2\pi\delta_{\alpha\beta}/N_\beta$
(${\alpha,\beta\in}\{x,y\}$).
We use $|\mathbf{k}\rangle$ and $|{k_x},{k_y}\rangle$ interchangeably.
The orbital $b$ is embedded at $\boldsymbol{\epsilon}_b$ relative to its unit 
cell coordinate in real space~\cite{Wu12:Wannier}.
The projected density in the Chern 
band is~\cite{Parameswaran12:W-inf,Goerbig12,Bernevig12:Counting}
\begin{equation}
\label{eq:density-chern}
\rho_\mathbf{q}=\sum_{\mathbf{k}}^\mathrm{BZ}
\left[\sum_b e^{-i\mathbf{q}\cdot\boldsymbol{\epsilon}_b}
u_b^*(\mathbf{k})u_b(\mathbf{k}+\mathbf{q})
\right]
|\mathbf{k}\rangle\langle{\mathbf{k}+\mathbf{q}}|,
\end{equation}
where $u_b(\mathbf{k})$ is the periodic part of the Bloch wave function.
At $\mathbf{q}=\mathbf{g}_\alpha$, the bracketed factor in 
Eq.~\eqref{eq:density-chern} gives the band geometry through the nonunitary 
exponentiated Abelian Berry connection,
$\mathcal{A}_\alpha=\sum_b e^{-i\mathbf{g}_\alpha\cdot\boldsymbol{\epsilon}_b}
u_b^*(\mathbf{k})u_b(\mathbf{k}+\mathbf{g}_\alpha)$.
$|\mathcal{A}_\alpha(\mathbf{k})|$ contains the quantum distance between 
$|\mathbf{k}\rangle$ and $|\mathbf{k}+\mathbf{g}_\alpha\rangle$, while
$A_\alpha(\mathbf{k})
=\mathcal{A}_\alpha(\mathbf{k})/|\mathcal{A}_\alpha(\mathbf{k})|$
is the unitary Berry connection between them.
We define $\rho_\alpha=\rho_{\mathbf{g}_\alpha}$.

The gauge-invariant Wilson loops (geometric phases) can be obtained by 
parallel transporting around a close loop over the BZ torus.
All the contractible loops consist of a product of loops around a single 
plaquette, namely
$\rho_x\rho_y[\rho_y\rho_x]^{-1}=
\sum_\mathbf{k}^{\mathrm{BZ}}
D(\mathbf{k})\, \Wsq(\mathbf{k})\,
|\mathbf{k}\rangle\langle\mathbf{k}|$.
Here, $D(\mathbf{k})=
|\mathcal{A}_x(\mathbf{k})\mathcal{A}_y(\mathbf{k}+\mathbf{g}_x)
\mathcal{A}^{-1}_x(\mathbf{k}+\mathbf{g}_y)
\mathcal{A}^{-1}_y(\mathbf{k})|\in\mathbb{R}$
is related to the nonuniformity of the 
quantum distance, and
$\Wsq(\mathbf{k})=A_x(\mathbf{k})A_y(\mathbf{k}+\mathbf{g}_x)
[A_y(\mathbf{k})A_x(\mathbf{k}+\mathbf{g}_y)]^\dagger\in\mathrm{U}(1)$
is the unitary Wilson loop around the plaquette with its lower-left corner at 
$\mathbf{k}$.
For large enough $N_x$ and $N_y$, we can unambiguously extract the Berry curvature 
$f_\mathbf{k}=\frac{1}{2\pi}\Im\log\Wsq(\mathbf{k})$, with 
finite-size normalization convention 
$\sum_\mathbf{k}^\mathrm{BZ}f_\mathbf{k}=C$.
$\Im$ takes the imaginary part in the principal branch $\Im\log(z)\in(-\pi,\pi]$.
This gives a sharp finite-size formula for the Chern number,
$C=\frac{1}{2\pi}\mathrm{Tr}\Im\log\!\left[
\rho_x\rho_y(\rho_y\rho_x)^{-1}
\right]$.
In addition to $\Wsq(\mathbf{k})$, there are also two independent
noncontractible Wilson loops on the torus, related to charge 
polarizations:
the Wilson loop around $k_y=0$, 
$W_x
=\mathrm{Phase}\left[\langle\mathbf{0}|\rho_x^{N_x}|\mathbf{0}\rangle\right]
=\langle N_x\mathbf{g}_x|\mathbf{0}\rangle
\prod_{\kappa=0}^{N_x-1}A_x(\kappa\mathbf{g}_x)$,
with $|\mathbf{0}\rangle\equiv|\mathbf{k}=\mathbf{0}\rangle$, 
and the Wilson loop $W_y$ around $k_x=0$ defined similarly.

The structure of geometric phases in the Chern band is fully specified by the 
collection of the Wilson loops $\Wsq(\mathbf{k})$ and $W_\alpha$, $\alpha=x,y$.
We now build a LLL basis in Bloch $\mathbf{k}$ space, from which all properties 
of a Chern band with arbitrary Chern number can be translated \emph{mutatis 
mutandis}.
Diagonalizing the Haldane pseudopotentials in this basis gives us the FCI 
model wave functions.

We consider electrons on a (continuum) torus 
$(\mathbf{L}_x,\mathbf{L}_y)\sim (N_x\mathbf{b}_x,N_y\mathbf{b}_y)$ with twist angle 
$\theta$ in a magnetic field $\mathbf{B}=B\hat{e}_z$.
The magnetic translations are $T(\mathbf{d})=e^{-i\mathbf{d}\cdot\mathbf{K}}$, where 
$\mathbf{K}=-i\hbar\nabla-e\mathbf{A}+e\mathbf{B}\times\mathbf{r}$.
We adopt the Landau gauge
$\mathbf{A}(\mathbf{r})=Bx\hat{e}_y$.
The guiding-center periodic boundary conditions $T(\mathbf{L}_\alpha)=1$ 
quantize the number of flux quanta $N_\phi=L_xL_y\sin\theta/(2\pi l_{\!B}^2)$ 
to an integer~\cite{Haldane85:TorusBZ}, where $l_{\!B}=\sqrt{\hbar/(eB)}$ is the magnetic length.
We set $N_\phi=N_xN_y$ in accordance with the Chern 
insulator~\cite{Qi11:Wavefunction,Bernevig12:Counting} for $C=1$.
The usual basis $\{|j\rangle\}$ in the LLL is
\begin{multline}
\label{eq:j-basis}
\langle x,y|j\rangle=
\frac{1}{(\sqrt{\pi}L_y l_{\!B})^{1/2}}
\sum_n^{\mathbb{Z}}
\exp\Big[
2\pi(j+nN_\phi)\frac{x+iy}{L_y}\\
-i\frac{\pi L_xe^{-i\theta}}{N_\phi L_y}(j+nN_\phi)^2
\Big]\,
e^{-x^2/(2l_{\!B}^2)}.
\end{multline}
To make contact with the Bloch states, we introduce a new LLL basis that 
diagonalizes translations in both directions,
$T(\mathbf{L}_\alpha/N_\alpha)|\mathbf{k}\rangle
=e^{-i2\pi k_\alpha/N_\alpha}|\mathbf{k}\rangle$,
\begin{equation}
\label{eq:k-basis}
|\mathbf{k}\rangle=\frac{1}{\sqrt{N_x}}\sum_{m=0}^{N_x-1}e^{i2\pi mk_x/N_x}
|j=mN_y+k_y\rangle,
\end{equation}
where $\mathbf{k}=\sum_\alpha k_\alpha\mathbf{g}_\alpha$ lives on the lattice 
reciprocal to $(\mathbf{L}_x,\mathbf{L}_y)$.
These states are periodic in $k_x$, 
$|{k_x+N_x},k_y\rangle=|k_x,k_y\rangle$, but
quasiperiodic~\footnote{The nonperiodicity signals a topological 
obstruction to a periodic smooth gauge (Chern number $C=1$) in the continuum 
limit.}
in $k_y$, $|k_x,{k_y+N_y}\rangle=e^{-i2\pi k_x/N_x}|k_x,k_y\rangle$. 
Each $|\mathbf{k}\rangle$ satisfies $T(\mathbf{L}_\alpha)=1$.
We find the LLL-projected density in the $|\mathbf{k}\rangle$ basis,
\begin{equation}
\label{eq:density-LLL}
\rho_\mathbf{q}=
e^{-\mathbf{q}^2l_{\!B}^2/4}
\sum_\mathbf{k}^\mathrm{BZ}
e^{-i2\pi q_x(k_y+q_y/2)/N_\phi}
|\mathbf{k}\rangle\langle\mathbf{k}+\mathbf{q}|,
\end{equation}
with $\mathbf{q}=\sum_\alpha q_\alpha\mathbf{g}_\alpha$, $q_\alpha\in\mathbb{Z}$.
The Wilson loops are $\Wsq(\mathbf{k})=e^{i2\pi/N_\phi}$, $W_x=e^{-i2\pi k_y/N_y}$, and 
$W_y=e^{i2\pi k_x/N_x}$.

Using Eq.~\eqref{eq:density-LLL}, one can diagonalize any FQH Hamiltonian 
$\sum_\mathbf{q}V_\mathbf{q}\rho_\mathbf{q}\rho_{-\mathbf{q}}$ (including 
pseudopotential and even higher-body Hamiltonians), directly in 
the $|\mathbf{k}\rangle$ basis, and then translate the resulting wave function 
to the FCI by replacing $|\mathbf{k}\rangle$ with the lattice Bloch states.
The advantage of the new LLL basis [Eq.~\eqref{eq:k-basis}] 
is many-fold. The conditions for the relevance of the FQH state to FCI are 
explicit in this basis [Eq.~\eqref{eq:density-LLL}]: The Berry curvature must 
not fluctuate wildly~\cite{Parameswaran12:W-inf} and the quantum 
distance~\footnote{To be precise, one minus the quantum distance as 
usually defined.} over the 
Chern band must fall off with 
$\mathbf{q}$ rapidly, similar to $e^{-\mathbf{q}^2l_{\!B}^2/4}$.
Equation~\eqref{eq:density-LLL} also allows a much simpler and more effective 
treatment of the curvature fluctuations in gauge fixing (see below).
The most practical advantage of working directly in Bloch basis is the 
avoidance of the many-body Fourier transform in the Wannier 
prescription. This greatly simplifies the numerical implementation and
nearly squares the largest Hilbert space dimension 
that we can study in numerics.

We now turn to the case of $C>1$ and construct a Bloch-like basis in the $C$-component 
LLL with $N_\phi=N_xN_y/C$ fluxes that forms an $N_x\times N_y$ BZ with flat 
curvature and Chern number $C$.
The starting point is to look for two commuting translation operators that 
resolve an $N_x\times N_y$ BZ.
The finite magnetic translations $T_\alpha= T(\mathbf{L}_\alpha/N_\alpha)$ seem 
natural, but they do not commute, $T_xT_y=T_yT_xe^{i2\pi/C}$.
The cure must come from the color structure of the multicomponent system.
We assume a color-neutral Hamiltonian $H$.
Two color operators $P$ and $Q$ (diagonal in real space) commute with the 
Hamiltonian,
\begin{equation}
\begin{aligned}
P|\sigma\rangle&=|\sigma+1\text{ (mod $C$)}\rangle,&
Q|\sigma\rangle&=e^{i2\pi\sigma/C}|\sigma\rangle.
\end{aligned}
\end{equation}
$|\sigma\rangle$, with $\sigma\in\mathbb{Z}_C$, are color eigenstates.
Their commutation relation $PQ=QPe^{-i2\pi/C}$ is complementary to that 
of $T_x,T_y$.
The two color-entangled operators $\widetilde{T}_x=T_xP$ and 
$\widetilde{T}_y=T_yQ$ commute with each other and with $H$~\footnote{
Alternatively, we can also use the operator pair $(T_xQ,T_yP^\dagger)$ to 
define the momentum eigenstates.
This amounts to substituting the color eigenstate $|\sigma\rangle$ in 
$|\mathbf{k}\rangle$ to 
$|t\rangle\equiv\frac{1}{\sqrt{C}}\sum_{\sigma=0}^{C-1}
e^{i2\pi t\sigma/C}|\sigma\rangle$,
For our purpose of obtaining a FCI model state, this change is just a trivial 
unitary transform that leaves the color-neutral Hamiltonian intact.}.
We define the eigenstates $|\mathbf{k}\rangle$ with
$\widetilde{T}_\alpha|\mathbf{k}\rangle=e^{-i2\pi k_\alpha/N_\alpha}|\mathbf{k}\rangle$,
\begin{multline}
\langle x,y,\sigma|\mathbf{k}\rangle
=\frac{1}{(\sqrt{\pi}N_xL_yl_B)^{1/2}}
\sum_n^{\mathbb{Z}}
e^{i2\pi (nC+\sigma)k_x/N_x}\\
\exp\Bigg[
2\pi\left(k_y+nN_y + \frac{\sigma}{C}N_y\right)\frac{x+iy}{L_y}\\
-i\frac{\pi L_xe^{-i\theta}}{N_\phi L_y}\left(k_y+nN_y + \frac{\sigma}{C}N_y\right)^2
\Bigg]\,
e^{-x^2/(2l_{\!B}^2)}.
\end{multline}
Because of $[T(\mathbf{L}_\alpha),\widetilde{T}_\beta]\neq 0$, generically we have 
to abandon the boundary condition $T(\mathbf{L}_\alpha)=1$ and adopt the 
color-entangled generalization ${\widetilde{T}_\alpha^{N_\alpha}=1}$, 
i.e.
\begin{equation}
\label{eq:color-entangled-PBC}
T(\mathbf{L}_x)P^{N_x}=T(\mathbf{L}_y)Q^{N_y}=1.
\end{equation}
This quantizes $k_\alpha$ to integers.
Since $\widetilde{T}_\alpha^{N_\alpha}$ commute with each other by 
construction, $N_\phi$ is not restricted to an integer any more, 
unlike~\cite{Barkeshli12:Nematic}.
We only require $N_x,N_y,C\in\mathbb{Z}$.
The $|\mathbf{k}\rangle$ states are periodic in $k_x$ but quasiperiodic in 
$k_y$, $|k_x,k_y+N_y\rangle=e^{-i2\pi k_xC/N_x}|k_x,k_y\rangle$.
There are $N_x\times N_y$ independent $|k_x,k_y\rangle$ states, which form a 
BZ of the same size as the lattice and with the same Chern number $C$.
After summing over colors, the LLL-projected density operator 
$\rho_\mathbf{q}=\sum_\sigma^C\rho_{\mathbf{q}\sigma}$ in the 
color-entangled basis $|\mathbf{k}\rangle$
takes identical form to Eq.~\ref{eq:density-LLL}, except for 
the generalization $N_\phi=N_xN_y/C$.
The color-entangled BZ has flat curvature $f_\mathbf{k}=1/N_\phi$, as inferred 
from $\Wsq(\mathbf{k})=e^{i2\pi/N_\phi}$.
The matrix elements of $\rho_\mathbf{q}$ in the $C$-component LLL,
which are the building blocks of the interacting Hamiltonian,
are exactly equal to the $C$th power of those in the single-component 
LLL.
Model wave functions of pseudopotential Hamiltonians in the $|\mathbf{k}\rangle$ 
basis can immediately be translated to the FCI with arbitrary $C$.
Further, we can generalize the color-entangled boundary conditions in the LLL to 
$\widetilde{T}_\alpha^{N_\alpha}=e^{-i2\pi\gamma_\alpha}$, where the twist angle 
$\gamma_\alpha\in\mathbb{R}$ corresponds to flux insertions.
This shifts the momentum $\mathbf{k}\rightarrow\mathbf{k}+\boldsymbol{\gamma}$ 
with $\boldsymbol{\gamma}=\sum_\alpha\gamma_\alpha\mathbf{g}_\alpha$.
The connections become $A_\alpha(\mathbf{k}+\boldsymbol{\gamma})$, while 
the large Wilson loops around $k_\alpha=0$ are 
$W_x(\gamma_y)=e^{-i2\pi C\gamma_y/N_y}$ and 
$W_y(\gamma_x)=e^{i2\pi C\gamma_x/N_x}$.

Linking together the LLL $|\mathbf{k}\rangle$ and the lattice 
$|\mathbf{k}\rangle$ bases requires one additional step of gauge fixing, 
$|\mathbf{k}\rangle\rightarrow e^{i\zeta_\mathbf{k}}|\mathbf{k}\rangle$.
After that, any many-body state $|\Psi\rangle_\mathrm{L}$ over
our colorful LLL can be transcribed to the FCI~\footnote{For actual lattice 
calculations, it is desirable to use periodic gauge with 
$|\mathbf{k}\rangle=|\mathbf{k}+N_\alpha\mathbf{g}_\alpha\rangle$ (no sum 
implied). Simply restricting $\mathbf{k}$ to a single BZ would achieve this, 
as long as the BZ choice for the lattice system is consistent with that for 
the LLL.},
\begin{equation}
\label{eq:bloch-construction}
|\Psi\rangle
=\sum_{\{\mathbf{k}\}}
e^{i\sum_\mathbf{k} \zeta_\mathbf{k}}
|\{\mathbf{k}\}\rangle\times
{}_\mathrm{L}^{\boldsymbol{\gamma}}\!\langle\{\mathbf{k}\}
|\Psi\rangle_\mathrm{L},
\end{equation}
where ${}_\mathrm{L}^{\boldsymbol{\gamma}}\!\langle\{\mathbf{k}\}|$ is the 
color-entangled occupation-number basis in the LLL with twist 
$\boldsymbol{\gamma}$.
See the Supplemental Material for the explicit construction of 
$e^{i\zeta_\mathbf{k}}$ and $\boldsymbol{\gamma}$. 

For FCI with $C>1$, previous studies suggested that the equivalent FQH 
states are the $\mathrm{SU}(C)$ color-singlet Halperin 
states~\cite{Barkeshli12:Nematic,Liu12:Higher,Sterdyniak12:Higher,Lu12:Parton}.
They are the exact zero modes of the color-neutral
LLL-projected Hamiltonian
$H_\mathrm{FQH}=\sum_\mathbf{q} V_\mathbf{q} \rho_\mathbf{q}\rho_{-\mathbf{q}}$,
where $\mathbf{q}$ is summed over the infinite lattice reciprocal to 
$(\mathbf{L}_x,\mathbf{L}_y)$ and
the interaction between color-neutral densities
$\rho_\mathbf{q}=\sum_\sigma\rho_{\mathbf{q}\sigma}$ is
$V_\mathbf{q}=V_0$ for bosons and 
$V_\mathbf{q}=V_0 + (1-\mathbf{q}^2l_{\!B}^2)V_1$ for 
fermions, with pseudopotential $V_n>0$~\footnote{We focus only on the 
color-singlet states as observed in numerics~\cite{Sterdyniak12:Higher}.}.
For the FQH effect in 2D electron gas, the boundary conditions 
$T(\mathbf{L}_\alpha)=1$ are imposed separately on different color components.
In the LLL description of a FCI, however, we require the system to be periodic 
under the color-entangled translations $\widetilde{T}_\alpha^{N_\alpha}$.
This breaks the $\mathrm{SU}(C)$ symmetry.
To compare with the Halperin $\mathrm{SU}(C)$-singlet states, we examine the 
commensurate case $N_x/C\in\mathbb{Z}$.
The boundary conditions in Eq.~\eqref{eq:color-entangled-PBC}
thread $\Phi_\sigma=\sigma N_y/C$ (color-dependent) magnetic fluxes 
along the $y$ direction into the
$\sigma$ component of the LLL~\footnote{We have verified by numerical 
diagonalization that the eigenstates of $H_\mathrm{FQH}$ with color-entangled 
boundary conditions indeed coincide with the usual Halperin states with 
$\Phi_\sigma$ flux insertion, when $N_x/C\in\mathbb{Z}$.}.
In the one-dimensional localized basis for the LLL [Eq.~\eqref{eq:j-basis}],
this shifts the Landau orbitals of color $\sigma$ by 
$\Phi_\sigma\mathbf{L}_x/N_\phi$ in real space.
Hence we propose that the Wannier mapping~\cite{Barkeshli12:Nematic} be 
modified to identify the hybrid Wannier states with our shifted LLL 
orbitals.
In the generic, noncommensurate case, the translation $T(\mathbf{L}_x)$ 
changes the color of the particle, due to $T(\mathbf{L}_x)P^{N_x}=1$.
Our construction thus provides a finite-size realization of the ``wormhole'' 
connecting different color components~\cite{Barkeshli12:Nematic}.

\begin{figure}[tb]
\centering
\includegraphics[]{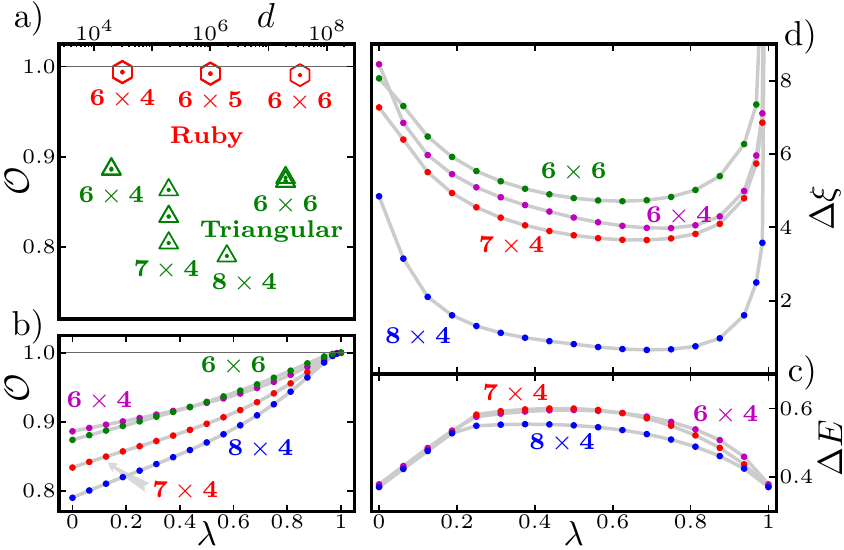}%
\caption{\label{fig:overlaps-adiabatic-continuity}
(a) shows the overlaps $\mathcal{O}$ between our FCI model states and the 
ground states of the fermionic ruby and the bosonic triangular lattice 
models, as a function of the Hilbert space dimension $d$.
(b)-(d) demonstrate the adiabatic continuity between the triangular 
lattice model and the color-entangled Halperin pseudopotential Hamiltonian on 
$6\times 4$, $7\times 4$, $8\times 4$, and $6\times 6$ lattices ($\nu=1/4$ filling).
We set $U=7.4237$, $7.0003$, $6.9677$, and $5.0955$, respectively, to equalize the energy gaps 
at $\lambda=0,1$.
(b) shows the overlaps $\mathcal{O}$ between our FCI model states and the 
ground states of the interpolation Hamiltonian $H_\lambda$.
(c) shows the energy gap $\Delta E$ above the ground states of $H_\lambda$. 
(d) shows the entanglement gap $\Delta \xi$ of the ground states of 
$H_\lambda$.
$\Delta\xi$ is defined as the gap between the low-lying structure identical to 
the full entanglement spectrum of the model states (at $\lambda=1$) and the 
higher levels. By this definition, $\Delta\xi$ is infinity at $\lambda=1$.
}
\end{figure}

We demonstrate the Bloch construction using the ruby lattice model
($C=1$)~\cite{Hu11:Ruby} and the two-orbital triangular lattice
model ($C=3$)~\cite{Yang12:Flatband}.
We construct the FCI model states through Eq.~\eqref{eq:bloch-construction}
from the exact-diagonalization ground states of $H_\mathrm{FQH}$ with 
color-entangled boundaries.
We find high overlaps [Fig.~\ref{fig:overlaps-adiabatic-continuity}(a)] and an
identical low-lying structure in the entanglement spectrum with the FCI 
ground states~\cite{Wu12:Zoology,Sterdyniak12:Higher}.
The 12-fermion Laughlin state on the ruby lattice model has a Hilbert 
space of dimension $3.4\times 10^7$.
This state is well captured by the model wave function obtained from our 
construction (overlap $\approx 0.99$).
The triangular lattice model has decent overlaps, albeit lower than the ruby 
lattice model.
The model we propose has the particle-hole symmetry, 
which is generally absent in the FCI 
models~\cite{Lauchli12:Hierarchy,Grushin12:Dispersion}.
When the lattice model exhibits such an emergent symmetry, our construction 
can also capture it.

To further examine our construction for $C>1$, we study the
interpolation Hamiltonian
$H_\lambda={(1-\lambda)}H_\mathrm{FCI}+\lambda H_\mathrm{FQH}$,
$0\le\lambda\le1$~\cite{Scaffidi12:Adiabatic,Liu13:Wannier}.
For bosonic on-site density-density interaction on the triangular lattice
$H_\mathrm{FCI}=
U\sum_{ab}\sum_{\{\mathbf{k}_{1-3}\}}
\widetilde{\psi}^\dagger_{\mathbf{k}_1a}
\widetilde{\psi}^\dagger_{\mathbf{k}_2b}
\widetilde{\psi}^{\phantom{\dagger}}_{\mathbf{k}_3b}
\widetilde{\psi}^{\phantom{\dagger}}_{\mathbf{k}_4a}$,
where $\mathbf{k}_4=\mathbf{k}_1+\mathbf{k}_2-\mathbf{k}_3$ (mod 
$N_\alpha\mathbf{g}_\alpha$), and
$\widetilde{\psi}^\dagger_{\mathbf{k}b}
=e^{i\zeta_\mathbf{k}}u^*_b(\mathbf{k})\psi^\dagger_\mathbf{k}$
is gauge fixed by $e^{i\zeta_\mathbf{k}}$, with 
$|\mathbf{k}\rangle=\psi^\dagger_\mathbf{k}|\emptyset\rangle$.
For $H_\mathrm{FQH}$, we use color-entanglement boundary conditions 
$\boldsymbol{\gamma}$.
We find that the FCI model states are adiabatically connected to the 
actual ground states: $H_\lambda$
remains gapped for $\lambda\in[0,1]$ and its ground states retain the 
characters of the FCI model states as seen in both overlaps and the particle 
entanglement spectrum [Fig.~\ref{fig:overlaps-adiabatic-continuity}(b)-(d)].
As observed in~\cite{Sterdyniak12:Higher}, the six-boson state on $6\times 4$ 
lattice has clear deviations from the usual Halperin state in the entanglement 
spectrum. Our FCI model state exactly reproduces these novel features.
Note that the $8\times 4$ lattice is closer to the thin-torus 
limit~\cite{Bernevig12:TT}, resulting in smaller overlaps and $\Delta\xi$ 
values.

In this Letter, we introduce a Bloch basis for multicomponent LLL with
a rational number of fluxes that entangles real and internal spaces on 
the one-body level.
We establish a Bloch-basis mapping between a Chern band 
with an arbitrary Chern number $C$ on an arbitrary
$N_x\times N_y$ lattice and a $C$-component LLL with 
$N_\phi=N_xN_y/C\in\mathbb{Q}$ fluxes.
This mapping leads to a novel scheme, which we call Bloch construction, to 
build FCI model states from color-neutral FQH Hamiltonians.
It treats bosonic/fermionic FCI with arbitrary $N_x,N_y,C\in\mathbb{Z}$ in a 
wholesale fashion,
and can handle large system sizes.
The new gauge fixing in our basis significantly improves the overlaps with the 
actual ground states when curvature strongly fluctuates.

We refer to the constructed FCI model states as the color-entangled 
Halperin states.
They are distinct from the $\mathrm{SU}(C)$-singlet Halperin states due to the 
color-entangled boundary conditions.
When the lattice size is commensurate with $C$, the color-entangled 
states are the generalization of the usual Halperin states to 
color-dependent twisted boundaries.
More generally, the lattice setup opens up access to the 
color-entangled, unphysical sectors of a multicomponent FQH system 
in a physical way.
Our new formalism can be applied to the NASS states, and can be used to 
extract the exclusion principle for the counting of low-lying levels in the 
energy and the entanglement spectra.

We wish to thank F.D.M.~Haldane, C.~Fang, and B.~Estienne for inspiring 
discussions, and thank A.~Sterdyniak and C.~Repellin for collaborations on 
related work.
BAB and NR were supported by NSF CAREER DMR-095242, ONR-N00014-11-1-0635, 
ARMY-245-6778, MURI-130-6082, Packard Foundation, and Keck grant.
YLW was supported by NSF CAREER DMR-095242.

\appendix

\section{Supplemental Material}

\subsection{Gauge Fixing}

The connections over the lowest Landau level (LLL) Brillouin zone (BZ) are 
$A^\mathrm{L}_x(\mathbf{k})=e^{-i2\pi k_y/N_\phi}$, and 
$A^\mathrm{L}_y(\mathbf{k})=1$ (superscript `L' represents LLL).
They satisfy the discrete analog of the Coulomb
gauge condition~\footnote{In the continuum limit~\cite{Wu12:Wannier}, the 
exponentiated connections become 
$A_\alpha(\mathbf{k})\approx
e^{i\mathbf{a}(\mathbf{k})\cdot \mathbf{g}_\alpha}$, where 
$\mathbf{a}(\mathbf{k})=-i\langle u_\mathbf{k}|\nabla_\mathbf{k}|u_\mathbf{k}\rangle$ is 
the Berry connection, with $|u_\mathbf{k}\rangle$ being the periodic part of 
the Bloch state.
The Coulomb gauge condition on $\mathbf{a}(\mathbf{k})$ is 
$\nabla_\mathbf{k}\cdot \mathbf{a}(\mathbf{k})=0$. This enables one to write 
the connection in terms of a stream function $\phi(\mathbf{k})$,
$\mathbf{a}(\mathbf{k})=\hat{e}_z\times\nabla_\mathbf{k}\phi(\mathbf{k})$.
Since $\nabla_\mathbf{k}\times \mathbf{a}(\mathbf{k})=F(\mathbf{k})\hat{e}_z$,
$\phi(\mathbf{k})$ satisfies a Poisson equation
$\nabla_\mathbf{k}^2\phi(\mathbf{k})=F(\mathbf{k})$, where $F(\mathbf{k})$ 
is the Berry curvature with the usual normalization 
$\int\mathrm{d}^2\mathbf{k}F(\mathbf{k})=2\pi C$.
},
i.e. they can be expressed in terms of a ``stream function'' 
$\phi^\mathrm{L}_\mathbf{k}=({k_y+1/2})^2/(2N_\phi)$ as
\begin{equation}
A^\mathrm{L}_\alpha(\mathbf{k})
=\exp\Big(-i2\pi\,\sum_\beta\varepsilon_{\alpha\beta}\,
[\mathrm{d}_\beta\phi^\mathrm{L}]_\mathbf{k}\Big).
\end{equation}
Here, $\mathrm{d}_\beta$ is the backward finite difference operator, defined by 
$[\mathrm{d}_\beta \phi]_\mathbf{k}=\phi_\mathbf{k}-\phi_{\mathbf{k}-\mathbf{g}_\beta}$,
and $\phi^\mathrm{L}_\mathbf{k}$ satisfies the discrete Poisson equation with 
curvature as source,
\begin{equation}
[\widetilde{\Delta}\phi^\mathrm{L}]_\mathbf{k}=1/N_\phi,
\end{equation}
with discrete Laplacian $\widetilde{\Delta}$ given by
\begin{equation}
[\widetilde{\Delta}\phi]_\mathbf{k}
=\sum_{\mathbf{p}}^{\pm \mathbf{g}_x,\pm \mathbf{g}_y}
\left(\phi_{\mathbf{k}+\mathbf{p}}-\phi_\mathbf{k}\right).
\end{equation}

We impose the same Coulomb gauge condition on the lattice connections,
and handle separately the average and the fluctuations of the lattice BZ curvature:
\begin{equation}
\label{eq:gauge-fixing-connections}
A^\mathrm{target}_\alpha(\mathbf{k})
=A^\mathrm{L}_\alpha(\mathbf{k}+\boldsymbol{\gamma})
\exp\left(-i2\pi\varepsilon_{\alpha\beta}[\mathrm{d}_\beta\phi]_\mathbf{k}\right).
\end{equation}
The non-zero curvature average necessitates the first factor above.
The shift $\boldsymbol{\gamma}=\sum_\alpha\gamma_\alpha\mathbf{g}_\alpha$ is 
determined by 
$W^\mathrm{lat}_x=W^\mathrm{L}_x(\gamma_y)$ and 
$W^\mathrm{lat}_y=W^\mathrm{L}_y(\gamma_x)$ (`lat' represents lattice), and it
accounts for the mismatch in the large Wilson loops between 
the two systems.
The curvature fluctuations are attended by the exponential factor, where the 
stream function $\phi_\mathbf{k}$ satisfies the discrete Poisson equation
$[\widetilde{\Delta}\phi]_\mathbf{k}=f_\mathbf{k}-1/N_\phi$, with boundary 
conditions 
$[\mathrm{d}_\alpha\phi]_\mathbf{k}=
[\mathrm{d}_\alpha\phi]_{\mathbf{k}-N_\beta\mathbf{g}_\beta}$
(no summation implied) and 
$\sum_\kappa^{N_x}[\mathrm{d}_y\phi]_{\kappa\mathbf{g}_x}=
\sum_\kappa^{N_y}[\mathrm{d}_x\phi]_{\kappa\mathbf{g}_y}=0$.
In plain words, we require that the connection corrections accounting for the 
curvature fluctuations should be periodic
over the lattice BZ~\footnote{The obstruction to simultaneous smoothness and 
periodicity is manifested in the non-fluctuating part 
$A^\mathrm{L}_\alpha(\mathbf{k}+\boldsymbol{\gamma})$.},
and they should not contribute to the large Wilson loops 
$W^\mathrm{lat}_\alpha$ which have already been fixed by the 
$A^\mathrm{L}_\alpha(\mathbf{k}+\boldsymbol{\gamma})$ factor.

Up to an inconsequential $\mathbf{k}$-independent constant, these conditions
allow a \emph{unique} solution 
\begin{equation}
\phi_\mathbf{k}=\varphi_\mathbf{k}+v_yk_x-v_xk_y,
\end{equation}
with $v_\alpha=\frac{1}{N_\alpha}\sum_{\kappa=0}^{N_\alpha-1}
\sum_\beta\varepsilon_{\alpha\beta}
[\mathrm{d}_\beta\varphi]_{\kappa\mathbf{g}_\alpha}$, and
\begin{multline}
\label{eq:periodic-stream-function}
\!\!\!\!
\varphi_\mathbf{k}=\frac{1}{N_xN_y}\sum_{\mathbf{n}\neq 0}
\frac{e^{i2\pi(k_xn_x/N_x+k_yn_y/N_y)}}
{2\cos(2\pi n_x/N_x)+2\cos(2\pi n_y/N_y)-4}\\
\sum_{\mathbf{p}}^\mathrm{BZ}
e^{-i2\pi(p_xn_x/N_x+p_yn_y/N_y)}
\left(f_{\mathbf{p}}-\frac{1}{N_\phi}\right),
\end{multline}
where $\mathbf{n}\equiv (n_x,n_y)$ runs over
$\{[0~..~N_x)\times[0~..~N_y)\}\backslash(0,0)$.

The connections $A^\mathrm{target}_\alpha(\mathbf{k})$ in 
Eq.~\eqref{eq:gauge-fixing-connections} are consistent with the actual 
(fluctuating) curvature over the lattice BZ.
Starting from a set of single-particle Bloch states $|\mathbf{k}\rangle$ with 
an arbitrarily chosen gauge and connections $A_\alpha(\mathbf{k})$,
our gauge fixing scheme amounts to the gauge transform 
$|\mathbf{k}\rangle\rightarrow e^{i\zeta_\mathbf{k}}|\mathbf{k}\rangle$ 
that reproduces $A^\mathrm{target}_\alpha(\mathbf{k})$,
\begin{equation}
e^{i\zeta_\mathbf{k}}=
\left[\prod_{\kappa=0}^{k_y-1}R_y(0,\kappa)\right]
\left[\prod_{\kappa=0}^{k_x-1}R_x(\kappa,k_y)\right],
\end{equation}
with $R_\alpha(\mathbf{k})=A^\mathrm{target}_\alpha(\mathbf{k})/A_\alpha(\mathbf{k})$~\footnote{
Despite the formal similarity of $e^{i\zeta_k}$ expressed as a product of 
ratios of connections, our gauge choice here is fundamentally different from 
the ``parallel-transport'' gauge~\cite{Wu12:Wannier} in the treatment of the 
curvature \emph{fluctuations}, embodied in the carefully constructed 
$A^\mathrm{lat}_\alpha(\mathbf{k})$.}.

\subsection{Emergent Particle-Hole Symmetry at Filling $\nu=2/3$}

As noted in the main text, the fractional quantum Hall system has 
particle-hole symmetry, which is absent in the fractional Chern 
insulators (FCI)~\cite{Lauchli12:Hierarchy,Grushin12:Dispersion}.
The anti-unitary particle-hole transformation $\mathcal{P}$ exchanges 
the band creation and annihilation operators 
$\psi_\mathbf{k}\leftrightarrow \psi^\dagger_\mathbf{k}$. The most generic 
normal-ordered two-body FCI Hamiltonian in the single-band approximation can 
be written as
\begin{equation}
H=\!\!\sum_{\{\mathbf{k}_{1-4}\}}^{\mathrm{BZ}} {\!\!\!\!}'\;
V_{\mathbf{k}_1\mathbf{k}_2\mathbf{k}_3\mathbf{k}_4}
\psi^\dagger_{\mathbf{k}_1}
\psi^{\phantom{\dagger}}_{\mathbf{k}_2}
\psi^\dagger_{\mathbf{k}_3}
\psi^{\phantom{\dagger}}_{\mathbf{k}_4}\!\!
-\sum_\mathbf{p}V_{\mathbf{k}\mathbf{p}\mathbf{p}\mathbf{k}}
\psi^\dagger_\mathbf{k}
\psi^{\phantom{\dagger}}_{\mathbf{k}},
\end{equation}
where the primed sum is constrained by
$\mathbf{k}_1+\mathbf{k}_3=\mathbf{k}_2+\mathbf{k}_4$ mod $\mathbf{g}_\alpha$, 
and the interaction coefficients satisfy
$V_{\mathbf{k}_1\mathbf{k}_2\mathbf{k}_3\mathbf{k}_4}^*
=V_{\mathbf{k}_2\mathbf{k}_1\mathbf{k}_4\mathbf{k}_3}$.
The particle-hole transformation changes the Hamiltonian by a one-body term 
plus a constant,
\begin{equation}
\mathcal{P}H\mathcal{P}^{-1}-H=
\sum_{\mathbf{k}}\varepsilon_\mathbf{k}
\psi^\dagger_\mathbf{k}
\psi^{\phantom{\dagger}}_{\mathbf{k}}
+\sum_{\mathbf{p}\mathbf{k}}^{\mathrm{BZ}}
V_{\mathbf{k}\mathbf{p}\mathbf{p}\mathbf{k}},
\end{equation}
where the effective dispersion $\varepsilon_\mathbf{k}$ is given by
\begin{equation}
\varepsilon_\mathbf{k}=\sum_\mathbf{p}
\Big(V_{\mathbf{p}\mathbf{k}\mathbf{k}\mathbf{p}}
+V_{\mathbf{k}\mathbf{p}\mathbf{p}\mathbf{k}}
-V_{\mathbf{k}\mathbf{k}\mathbf{p}\mathbf{p}}
-V_{\mathbf{p}\mathbf{p}\mathbf{k}\mathbf{k}}\Big).
\end{equation}
We emphasize that $\varepsilon_\mathbf{k}$ comes from the interaction and is 
unrelated to the single-particle dispersion of the Bloch band.
In general, $\varepsilon_\mathbf{k}$ has a non-trivial $\mathbf{k}$ 
dependence, and this breaks the particle-hole symmetry of the lattice model.
We can apply our construction of FCI model wave functions to test the possible 
presence of emergent particle-hole symmetry in the lattice models that support 
a Laughlin-like state.
We examine the ruby~\cite{Hu11:Ruby} and the kagome~\cite{Tang11:Kagome} 
lattice models with Chern number $C=1$. We focus on the $\nu=2/3$ filling 
factor, where the particle-hole conjugate of the $\nu=1/3$ Laughlin state 
should appear.

\begin{figure}[tb]
\centering
\includegraphics[]{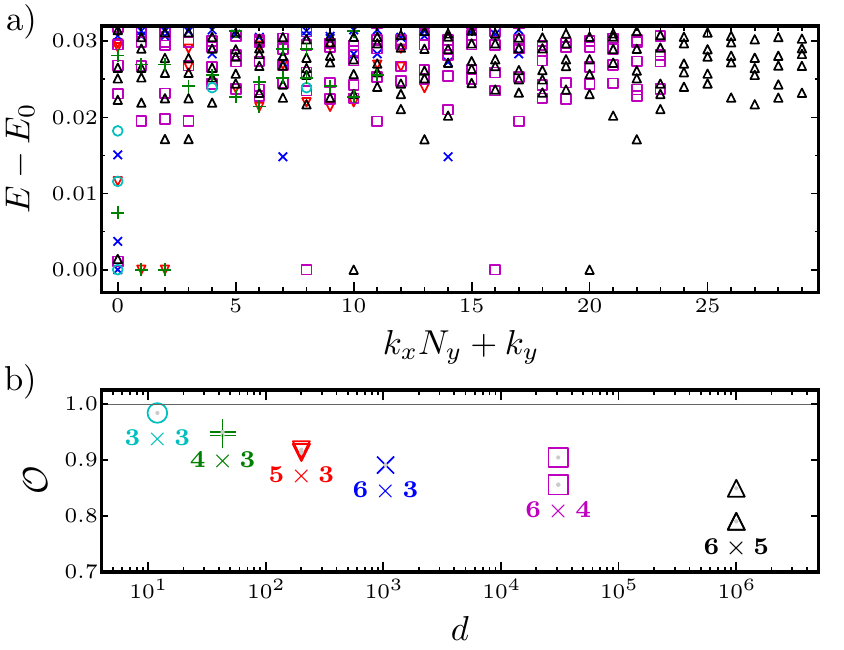}%
\caption{\label{fig:ruby-laughlin-two-thirds}
Panel a) shows the low energy spectrum of the ruby lattice model at filling 
$\nu=2/3$ for various system sizes, with energies shifted by $E_0$, the lowest 
energy for each system size.
Panel b) shows the overlaps $\mathcal{O}$ between our FCI $\nu=2/3$ Laughlin 
states and the lowest energy states in the corresponding momentum sectors of 
the ruby lattice model for various system sizes.
The system size $(N_x,N_y)$ represented by each group of markers is annotated 
in panel b).
}
\end{figure}

For the ruby lattice model, we observe gapped three-fold ground state in the 
energy spectrum, as shown in Fig.~\ref{fig:ruby-laughlin-two-thirds}a).
A clear energy gap above the three-fold ground state is visible when the 
number of particles is higher than $12$.
Using the formalism detailed in the main text, we construct the FCI Laughlin 
state at filling $\nu=2/3$. We find reasonable overlaps between these model 
states and the ground states of the ruby lattice model, as shown in 
Fig.~\ref{fig:ruby-laughlin-two-thirds}b). Compared with the conjugate states 
at filling $\nu=1/3$, the overlap values here are considerably smaller.
For the kagome lattice model, we do not observe gapped ground states in the 
energy spectrum. This model does not exhibit any trace of the particle-hole 
conjugate Laughlin state.
We note that the presence of a robust $\nu=1/3$ Laughlin state in a Chern 
insulator does not guarantee the existence of its particle-hole conjugate at 
$\nu=2/3$.

\end{document}